# Throughput Optimization of Coexistent LTE-U and WiFi in Next Generation Networks

Suganya S/Associate Professor
Dept. of ECE
CMRIT, Bangalore
India
suganya.senthil2005@gmail.com

Dr.Ramesh C/Professor
Dept. of ECE
CMRIT, Bangalore
India
crameshmail@gmail.com

Prof. Sumit Maheshwari
Dept. of ECE
CMRIT, Bangalore
India
sumitece87@gmail.com

*Abstract*—Next generation networks are envisioned to have ubiquitous availability and seamless access as main goals. In general, coexistence of multiple access technologies is one of the most promising way to achieve these goals, particularly using WiFi and LTE (Long Term Evolution) simultaneously which are two major players in future Internet. LTE is primarily used in a licensed spectrum but an unlicensed version is also in use by research communities. In this work we study LTE-U (LTE-Unlicensed) and WiFi Coexistence in a handheld device by analyzing existing scheduling mechanisms and proposing a novel scheme of cooperation. Throughput and fairness of these multi-homed technologies are studied in detailed. Also, we propose to use a deferral based proactive scheme for improved system utilization.

*Keywords*—LTE-U, WiFi, Scheduling, Resource Blocks, Proportional Fair.

## I. Introduction

In recent years, coexistent of LTE-U (Long Term Evolution Unlicensed), which is the unlicensed band of LTE, and WiFi, which is another unlicensed under the ISM (Industry Science and Medical) band, is popular among the researchers as well as industries. Recent studies have provided multiple research challenges for coexistence of WiFi and LTE access technologies being accessed by a single device. On one hand, LTE is a resource hungry technology which suppresses the usage of WiFi while on the other hand WiFi is a cooperative access mechanism with methods such as Carrier Sense Multiple Access – Collision Avoidance (CSMA/CA) and Request to Send/Clear to Send (RTS/CTS) to avoid contention based collision.

In [1], a resource scheduling algorithm using inverse weights is proposed for LTE by considering proportional fair as well weighted fair queuing schemes. Proportional fair algorithms are very critical in joint optimization cases and a similar one is explored in [2] which provides an analytical model to calculate the probability of collision of two networks. On similar lines, a proportional fairness based scheme is also detailed in [3] where the authors considered LTE-U and WiFi networks and their states are modeled using a multi-state Markov model where MAC (Medium Access Control) and physical layer cross optimization is detailed.

A multi-RAT (radio access technology) based model is used in [4] which provides fairness by using traditional ALOHA based technique for quick scheduling and a min-max based approach for global optimization. This work also includes a proportional fair mechanism to use retention probability as well as coverage probability for the performance metrics.

MAC layer approaches are popular as well in research community where the frames are altered, timings are changed or access mechanisms are modified in order to provide a smooth inter-network transition, unified access control and interoperability. LTE blank subframe allocation based method is proposed in [5] which imitate the technique of TV signal whitespaces. As LTE is an active first technology which tries to schedule all the resources as soon as possible while WiFi is a dormant technology employing carrier sensing based deferral which hampers itself.

As LTE release 13 considers deployments in unlicensed bands, [6-7] provides a detailed analysis of its coexistence with WiFi. A Markov model based listen before talk scheme is proposed to show that WiFi performance can be greatly improved [6]. Also, to solve the feasibility problem of coexistence, a study is reported in [8]. The variability of latency for a fixed sized packets is reported in [9] which encourages a need of generalized system to capture the effects of network parameters to optimize performance.

The pervasive challenges in this field are not only limited to cooperation but also are manifold including but not limited to (a) throughput maximization, (b) system optimization (c) energy efficiency (d) delay minimization, and (e) fairness. All these issues can be viewed as disparate set of requirements but are inter-related wherein solving one will mean to affecting other. In this work, we explore the opportunities of proactively scheduling LTE-U and WiFi resources using a deferral based scheme which not only is shown to optimize system throughput but also provides better system utilization.

The rest of the paper is organized as follows. Section II outlines the scheduling mechanisms in LTE and WiFi in detail. Proposed scheduling mechanism is explained in Section III. Section IV covers the system set-up and analysis framework. Results and analysis are presented in Section V while Section VI provides concluding remarks and outlining future work in the field.

## II. SCHEDULING IN LTE AND WIFI

In a coexistent scenario as shown in Fig. 1, cooperation is highly important. In present networks, there is minimal or no cooperation available between inter-network technologies which poses critical question on how to schedule the traffic and nodes in such a heterogeneous environment.

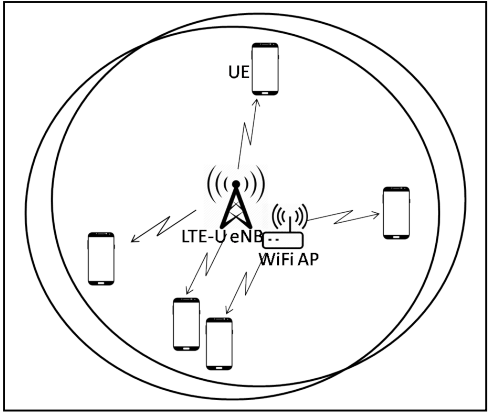

Figure 1: LTE-U and WiFi Coexistence Scenario

As it is well known that industry deployed LTE till date uses proportional fair queuing which tries to meet minimum throughput requirements of each of the UEs in the network while WiFi uses RTS-CTS based scheduling mechanism which is not as proactive as LTE. The clear demarcation between these two technologies is the way the channel is assigned to a user. On one hand former uses a semi-persistent scheduler which perpetually keeps track of all the available resources while later uses a simple availability based scheduling without any priority implementation.

Proportional fair scheduling can be described as equation (1) where the past average throughput is set as a weighing factor for the rate which is expected.

$$m_{i,k}^{PF} = d_k^i(t) / \bar{R}^i(t-1) \tag{1}$$

Clearly, a user whose past average throughput (denominator) is low, will get higher *m* value and therefore will be scheduled where *d* is the current throughput value (numerator). Also, a user with good channel condition will be scheduled too in which case the numerator will be high itself. This method provides fairness but lacks the overall system throughput optimization.

Augmenting this with WiFi has inherent advantages such as filling the gaps of spectrum and providing better throughput. WiFi uses the binary exponential back-off mechanism where the user is backed off in a time slot chosen with probability *p* in (0-$2^k$-1) where k is the number of successive collision for a user. Clearly, we can observe that while LTE provides a channel-aware scheduling, WiFi focuses on collision-aware scheduling which defers a user to multiple time slots before it can transmit (this is based on RTS collisions as described earlier).

Therefore, we observe two major challenges here with the combine LTE and WiFi scheduling, first there is a need of cooperation between the two and second, shortcoming of each should be complemented by the other. This can be achieved at control and data plane level as shown in Figure 2. The global network view has the centralized information about both the networks while LTE and WiFi controller exchange the information among each other which are scheduling, channel and UE parameters such as location, speed, battery level and so on.

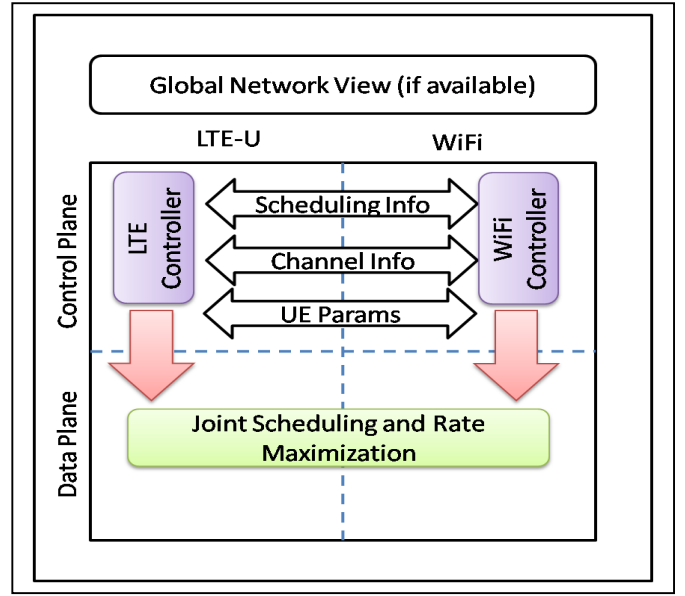

Figure 2: Proposed Network for Joint Scheduling

It is to be noted that the global view is not always available and therefore cooperation is crucial. Assuming cooperation between WiFi access point and LTE-U base station sharing scheduling information, channel information and user equipment (UE) parameters such as speed, direction etc., we propose a deferral based joint scheduling which is described in the next section.

## III. DEFERRAL BASED JOINT SCHEDULING

Consider the scenario where a UE has multiple interfaces in general. In precise, it has an LTE-U interface and a WiFi interface. In a multi-user scenario, some devices might not have capability of LTE (for example a laptop or a tablet) while others might not have WiFi capability (for example a feature phone with no WiFi). The objective of our scheduling is to provide a fair and globally optimal resource share to each of the device present in the network.

Let us assume that the LTE can provide rate $0 < R_L < R_{LMAX}$ to a user and WiFi can support rate $0 < R_W < R_{WMAX}$ to a user. Assuming system has K LTE only users, M WiFi only users and N users with both the interfaces active amounting to total K+M+N users.

Therefore, the objective of our work is to:

$$\text{maximize} \sum_{1}^{K+M+N} R_i$$

subject to:

$$R_{min} < R_i, i = 1, 2, .., K + M + N$$
$$R_j < R_{LMAX}, j = 1, 2, .., K$$
$$R_t < R_{WMAX}, t = 1, 2, .., M$$
$$R_p < R_{LMAX+WMAX}, p = 1, 2, .., N$$

As we can see that this is a multi-criteria optimization problem, we can heuristically solve using the cooperative networks. Also, we can see that the first constraint provides the fairness condition where each user is guaranteed a minimum rate which is very crucial for long term sustainability of a system.

Since this is a global optimization problem, it requires readjusting rates at each time instant for all the users which is an impractical solution therefore we provide an additional deferral block to the solution which works as shown in Figure 3. In this method, the countMax is the maximum number of times a UE goes below the minimum threshold before it is retrained at global optimizer along with similar UEs which face the same problem. This method avoids the global trainer to train complete system at all time instants and therefore increases stability along with fairness. Also, it is not always possible to get global optimal result for all the users in a network due to system limitations.

## IV. RESULTS AND DISCUSSION

We simulated the proposed model along with the comparison cases and the results are as follows. Figure 4, LTE base throughput without using WiFi interface for a group of users. Figure 5 illustrates the WiFi throughput without using LTE interface. The throughput depends upon the position of a user in the cell and the channel conditions.

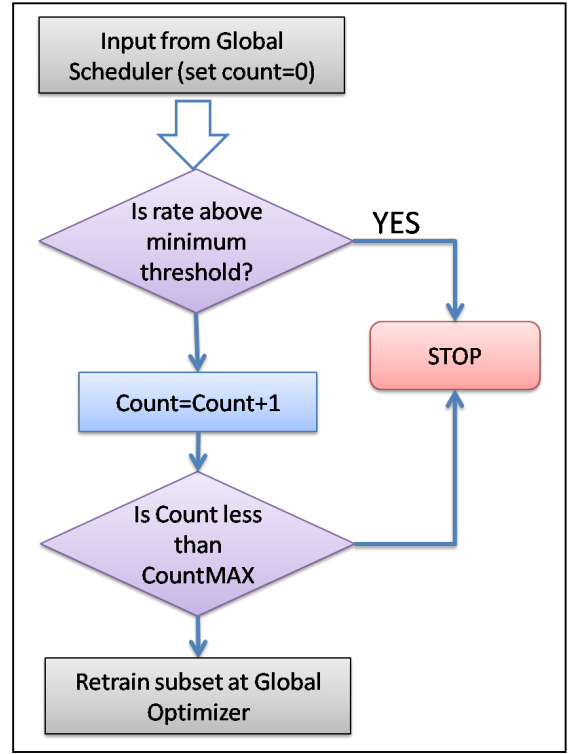

Figure 3: Deferral Based Joint Scheduler

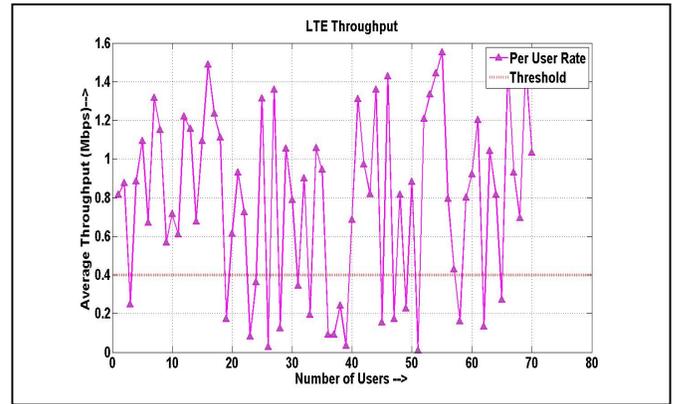

Figure 4: LTE Throughput without Joint Scheduler

As shown in Figure 6, the joint system optimization of LTE and WiFi provides a much higher throughput. The effect is due to the fact that the complete system is treated in a uniformly fair manner wherein the UE which does not have LTE interface is also provided enough bandwidth to sustain even the modern so called killer applications. The novelty of our work lies in the regime of deferral based optimization which reduces the load on a global optimizer. Also, the sharing of information among the access points provides an additional advantage which is yet to be explored in detail. Overall contribution of our work is to propose and showcase a novel system and scheme for the next generation of schedulers.

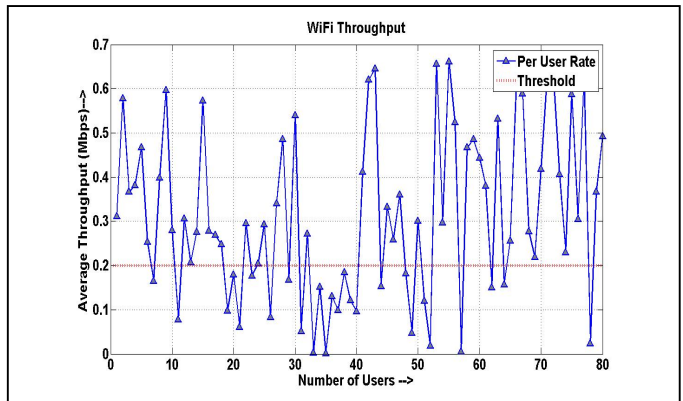

Figure 5: WiFi Throughput without Joint Scheduler

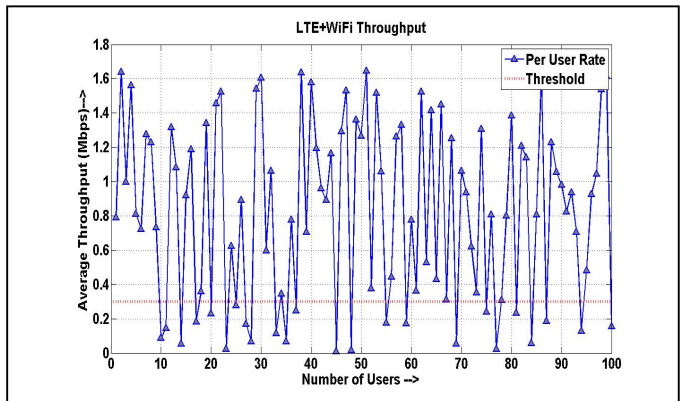

Figure 6: Throughput for WiFi and LTE-U Joint Scheduling

Table I represents the various statistics of the scheduling. It can be easily observed that the system allows some users to go below the threshold but the same users are recovered after countMax time interval. This method provides system an opportunity for optimization of throughput when SNR (Signal to Noise Ratio) for a user is higher. The LTE with WiFi improves the system throughput by 15% as compared to the LTE alone. When comparing with WiFi alone, the system throughput doubles owing to high data rate of LTE. The users who are below the threshold are just a little more than LTE and much lower than those served by WiFi alone without LTE. In general, the system is improving fairness and throughput by providing similar opportunities of access.

It is also observed that the standard deviation is minimum for the joint scheduler implying that no two users are far apart in throughput and therefore the system is fair for all of them.

TABLE I: Scheduler Statistics

|  | LTE | WiFi | LTE+WiFi |
|---|---|---|---|
| Users below Threshold | 20 | 28 | 22 |
| System Throughput (Mbps) | 54.47 | 28.87 | 62.26 |
| Max Throughput (Mbps) | 1.49 | 0.68 | 1.65 |
| Min Throughput (Mbps) | 0.003 | 0.003 | 0.02 |
| Standard Deviation | 0.53 | 0.6 | 0.48 |

V. CONCLUSION AND FUTURE WORK

Seamlessness is a topmost priority in the next generation networks where anytime, anywhere and any device is the main goal. In general, coexistence of multiple access technologies is a way to achieve these goals, particularly using WiFi and LTE simultaneously. LTE is primarily used in a licensed spectrum but an unlicensed version is also popularly in research communities. In this work we study LTE-U and WiFi coexistence in a handheld device by analyzing existing scheduling mechanisms and proposing a novel scheme of cooperation. Throughput and fairness of these multi-homed technologies are also studied in detail. Proposing a deferral based proactive scheme for improved system utilization, we found that LTE and WiFi together can provide a system throughput around 15% more than LTE alone and double of WiFi alone. The cooperation also improves fairness and throughput standard deviation reduces by 10%. Our future work will focus on optimizing the model and training using prediction based approach.

ACKNOWLEDGMENT

We appreciate the help of CMR Institute of Technology for supporting the work throughout.